\documentclass[
]{ceurart}

\usepackage{listings}
\usepackage{url}
\begin{document}

\copyrightyear{2026}
\copyrightclause{Copyright for this paper by its authors.
  Use permitted under Creative Commons License Attribution 4.0
  International (CC BY 4.0).}

\conference{ROMCIR 2026: The 6th Workshop on Reducing Online Misinformation through Credible Information Retrieval (held as part of ECIR 2026: the 48th European Conference on Information Retrieval), April 2, 2026, Delft, The Netherlands}

\title{Overview of ROMCIR 2026: The 6th Workshop on Reducing Online Misinformation through Credible Information Retrieval}


\author[1]{Marcos~Fernández-Pichel}[%
orcid=0000-0002-6560-9832,
email=marcosfernandez.pichel@usc.es,
url=https://citius.gal/team/marcos-fernandez-pichel/,
]
\address[1]{University of Santiago, Centro Singular de Investigación en Tecnoloxías Intelixentes (CiTIUS), Rúa de Jenaro de la Fuente Domínguez -- 15782 Santiago de Compostela, Spain}

\author[2,3]{Marinella Petrocchi}[%
orcid=0000-0003-0591-877X,
email=marinella.petrocchi@iit.cnr.it,
url=https://www.iit.cnr.it/en/marinella.petrocchi,
]
\address[2]{Institute of Informatics and Telematics -- CNR, Via G. Moruzzi, 1 -- 56124 Pisa, Italy}
\address[3]{IMT School for Advanced Studies, Piazza San Francesco, 19 – 55100 Lucca, Italy}

\author[4]{Kevin Roitero}[%
orcid=0000-0002-9191-3280,
email=kevin.roitero@uniud.it,
url=https://kevinroitero.com/,
]
\address[4]{University of Udine, Department of Mathematics, Computer Science and Physics, Via delle Scienze, 206 -- 33100 Udine, Italy}

\author[5]{Marco Viviani}[%
orcid=0000-0002-2274-9050,
email=marco.viviani@unimib.it,
url=http://www.ir.disco.unimib.it/people/marco-viviani/,
]
\cormark[1]
\address[5]{University of Milano-Bicocca, Department of Informatics, Systems, anc Communication (DISCo), Edificio U14 (ABACUS), Viale Sarca, 336 – 20126 Milan, Italy}

\cortext[1]{Corresponding author.}

\begin{abstract}
In the digital online ecosystem, we are surrounded by distinct forms of information pollution, posing significant threats to both individuals and society. Fake news, for instance, wields power to sway public opinion on matters of politics and finance. Deceptive reviews can either bolster or tarnish the reputation of businesses, while unverified medical advice may steer people toward harmful health practices. 
In light of this challenging landscape, it has become imperative to ensure that users have access to both topically relevant and factually accurate information that does not warp their perception of reality, and there has been a surge of interest in various strategies to combat misinformation through different
contexts and multiple tasks.
The purpose of the ROMCIR Workshop, for some years now, is precisely that of engaging the Information Retrieval community to explore potential solutions that extend beyond conventional misinformation detection approaches. Key objectives include identifying subjective and objective factors associated with information credibility and truthfulness, respectively, and integrating such factors as fundamental dimensions of relevance within IR Systems (IRSs), achieving early detection of misinformation, and ensuring that the search results retrieved are not only truthful but also explainable to the users of IRSs. Moreover, it is essential to evaluate the role of generative models such as Large Language Models (LLMs) in inadvertently amplifying misinformation problems, and how they can be used to support IRSs, together with the contribution that the human-in-the-loop paradigm can have in this context.
\end{abstract}


\maketitle

\section{Introduction}

The sixth edition of the ROMCIR Workshop concerns providing access not only to topically relevant information, but also to truthful/reliable/factually accurate information \textcolor{black}{in the online digital ecosystem}, to mitigate both the human-generated and AI-generated information disorder phenomenon with respect to distinct domains. By “information disorder” we mean all forms of communication pollution, from misinformation made out of ignorance, \textcolor{black}{automatically built on the basis of biased content,} to intentional sharing of false content \textcolor{black}{(generated both manually and automatically)} \cite{wardle2017information}.
Tackling information disorder is inherently complex. It involves the analysis of heterogeneous content types, dissemination platforms, and user intentions. This challenge is further intensified by structural features of the digital environment, such as \textit{filter bubbles} and \textit{echo chambers}, which reinforce users’ pre-existing beliefs and limit exposure to diverse perspectives \citep{bozdag2015breaking,del2016spreading,villa2021echo,bowties2023,DBLP:journals/corr/abs-2308-01750,impicciche2024comparing}. Emerging AI-related concerns further complicate this landscape. These include the \textit{explainability} of search results \citep{anand2023explainable,upadhyay2023explainable,upadhyay2025enhancing}, the evaluation of truthfulness in \textit{User-Generated Content} (UGC) \citep{viviani2017credibility,saridou2020misinformation,soprano2021many}, the potential of \textit{crowdsourcing} as a verification tool \citep{roitero2023can,la2024crowdsourced}, and the responsible integration of \textit{generative models} within IR systems \citep{cabitza2022responsible,najork2023generative}.
Another critical dimension is the protection of \textit{data confidentiality}, particularly in unstructured data contexts \citep{livraga2019data,livraga2023unveiling} and framed within IR applications \citep{wu2020dummy,cassani2024assessing}, where LLMs may play a transformative yet delicate role \citep{herranz2025can}. In light of these developments, designing effective and reliable experimental evaluation paradigms for IR systems becomes not only necessary but foundational to progress in this domain \citep{lioma2017evaluation,Suominen21,barron2024clef}.

\section{Aim and Topics of Interest}

Within the ECIR conference, the ROMCIR Workshop addresses a broad range of topics related to online misinformation/disinformation and trustworthy Information Retrieval. These topics span various types of content (e.g., Web pages, news articles, reviews, medical data), platforms (e.g., social media, microblogs, Q\&A systems), and user goals (e.g., detecting falsehoods, retrieving truthful information). In addition, the Workshop engages with emerging AI-related challenges, such as the explainability of search results, the evaluation of truthfulness in AI-generated content, and the integration of generative models to enhance \textit{Information Retrieval Systems} (IRSs). Accordingly, the topics of interest for ROMCIR 2026 include, but are not limited to:

\begin{itemize}
    \item Access to and retrieval of truthful/reliable information;
    \item Bot/spam/troll detection;
    \item Computational fact-checking;
    \item Credibility assessment of online documents;
    \item Crowdsourcing for information truthfulness/reliability assessment;
    \item Disinformation/misinformation/bias detection;
    \item Evaluation strategies to assess information truthfulness/reliability;
    \item Generative models and information truthfulness/reliability assessment;
    \item Human-in-the-loop misinformation detection;
    \item Information polarization in online communities, echo chambers;
    \item Propaganda identification/analysis;
    \item Query reformulation strategies for truthful/reliable IR;
    \item Search patterns and simulation strategies for information verification;
    \item Security, privacy, and information truthfulness/reliability;
    \item Societal reaction to misinformation;
    \item Trust and reputation.
\end{itemize}

\section{Keynote Speech}

Also this year, ROMCIR hosted a keynote speech by an expert in the field, namely Tommaso Caselli, who discussed recent advances in misinformation detection, outlining key challenges and its broader societal impact.

\paragraph{Misinformation Detection: Progress, Pitfalls, and the Path to Societal Impact.}

Over the past decade, expressions like “fake news”, “misinformation”, and “alternative facts” have increasingly entered everyday discourse, accompanied by a surge of research interest in the automatic detection of “incorrect information”. Two major events, the 2016 US Presidential Election and the COVID-19 pandemic, accelerated this trend, spurring a wave of research into detection tools and benchmark datasets. Yet, despite this momentum, a troubling gap has emerged between technical progress and societal impact. Drawing on my own work in this field, this talk critically examines the state of the art in misinformation detection: what has been accomplished, what genuinely works, and what remains unresolved. The field tends to prioritize incremental improvements over the needs of those on the front lines of the misinformation crisis: journalists, fact-checkers, and the broader public. Moreover, a fundamental challenge remains largely unaddressed: how do we actually help people correct false beliefs once they have taken hold? This talk is an invitation to reflect, and potentially to reorient. If misinformation detection is to deliver on its promise, we, as a community, must engage more deeply with real-world practitioners and start measuring success on outcomes rather than benchmarks.

\paragraph{Tommaso Caselli.} 
\begin{wrapfigure}{l}{0.213\textwidth}
\vspace{-13pt}
    \includegraphics[width=0.213\textwidth]{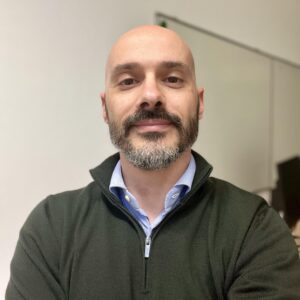}
      \vspace{-22pt}
\end{wrapfigure} He is a Senior Assistant Professor at the Center for Language and Cognition Groningen (CLCG), Faculty of Arts, University of Groningen (\url{https://www.rug.nl/staff/t.caselli/}). 
His research focuses on Natural Language Processing, in particular event extraction and framing, as well as the detection and mitigation of hate speech and misinformation.
He is a co-founder of the Events and Stories in the News workshop series and co-editor of the volume Computational Analysis of Storylines (Cambridge University Press, 2021). He has organized several semantic evaluation campaigns for English, Dutch, and Italian, and has contributed to the development of linguistic resources for these languages.
His work appears in major computational linguistics conferences and journals, and has received two Outstanding Paper Awards (COLING 2022 and ACL 2023). From 2023 to 2025, he served as coordinator of the “AI and Language” theme within the Jantina Tammes School of Digital Society, Technology, and AI at the University of Groningen.

\section{Submissions}

The ROMCIR 2026 Workshop received 10 submissions, of which 5 were accepted, resulting in an acceptance rate of 50\%. The authors of the accepted submissions were affiliated with universities from six different countries, including France, Italy, Spain, Turkey, United Kingdom, and Vietnam. This year's submissions particularly focused on the issues of: $(i)$ Explainable and Multimodal Fact-Checking Methods, $(ii)$ Reliability and Evaluation of Generative Models, and $(iii)$ Datasets and Human-Centered Evaluation for Misinformation Detection \citep{pichel2026romcir}.

Concerning research issue $(i)$, two \textit{full papers} were accepted. The two works tackle multimodal fact-checking on how to build verification systems and improve their technical capabilities. In particular, \citet{explainablefact} provide a comprehensive overview of explainable Automated Fact-Checking (AFC), addressing the transparency gap in current AI-driven misinformation detection. The proposed solution synthesizes conventional methods alongside recent advances utilizing Large Language Models (LLMs) and Vision-Language Models (VLMs). The study also reviews state-of-the-art datasets and evaluation metrics to offer a broader understanding of available resources. Finally, it highlights current limitations and proposes research directions to enhance the trustworthiness and transparency of AFC systems. \citet{imfact} introduce the first benchmark dataset specifically designed for near-duplicate image retrieval within the context of automated fact-checking. The dataset comprises 2,473 images organized into clusters of near-duplicates, each accompanied by professional claim reviews. The authors explore a contrastive learning framework to fine-tune vision models, simulating realistic image modifications to enhance retrieval performance. The study demonstrates that specialized training and benchmarking resources can significantly strengthen the detection and verification of visual misinformation.

Concerning research issue $(ii)$, a \textit{full} and a \textit{short paper} were accepted, both bringing together research on LLMs  robustness, hallucinations, and behavior under uncertainty or noise. In particular, \citet{ragreliability} introduces an evaluation protocol to systematically test the reliability of Retrieval-Augmented Generation (RAG) systems in misinformation-rich environments. This work specifically examines how models handle conflicts between their internal parametric knowledge and external evidence that may be clean, poisoned, or mixed. Using metrics like  ``Parametric Override Rate'' and ``Confidence Inflation'', the study identifies ``breaking points'' where misleading evidence begins to dominate model decisions. The findings offer quantitative insights into the vulnerabilities of RAG systems when exposed to realistic, unreliable retrieval contexts. \citet{benchmarksllm} investigate whether existing benchmarks underestimate the performance of LLMs in detecting contextual hallucinations. The authors propose an adjudication-based evaluation framework that compares original benchmark annotations with LLM-generated judgments re-evaluated by human adjudicators. Their analysis reveals that LLMs can capture subtle factual inconsistencies often overlooked during single-pass human annotation. Consequently, the paper suggests that model-assisted re-evaluation is necessary to create more reliable and semantically grounded benchmarks for ambiguity-prone tasks.

Concerning research issue $(iii)$, a \textit{full-paper} was accepted. \citet{deepfakedetection} evaluate the effectiveness of crowdsourcing for the detection of audiovisual deepfakes across two major benchmarks. The study measures how consistently non-experts can distinguish authentic from manipulated videos and identify specific manipulation types, such as audio-only or joint audio-video edits. Results indicate that while crowd judgments provide a scalable and useful screening signal, reliable modality attribution remains a significant challenge. The authors conclude that crowdsourcing is most effective as a preliminary stage in broader, multi-step verification workflows.

\section{Past Editions}

The first five editions of the ROMCIR Workshop, all co-located with the ECIR conference, led to fervent discussion and presentation of innovative work concerning a variety of open issues related to information disorder and IR. The first edition 
took place in online mode on April 1, 2021.\footnote{\url{https://romcir.disco.unimib.it/2021-edition/2021-workshop/}} 
The second edition 
took place both in person in Stavanger, Norway, and online, on April 10, 2022.\footnote{\url{https://romcir.disco.unimib.it/2022-edition/2022-workshop/}} 
The third edition 
took place in person in Dublin, Ireland, on April 2, 2023.\footnote{\url{https://romcir.disco.unimib.it/2023-edition/2023-workshop/}} The fourth edition took place in person in Glasgow, Scotland, on March 24, 2024.\footnote{\url{https://romcir.disco.unimib.it/2024-edition/2024-workshop/}} The fifth edition took place in person in Lucca, Italy, on April 10, 2025.\footnote{\url{https://romcir.disco.unimib.it/2025-edition/2025-workshop/}} 
The papers accepted at ROMCIR in its various editions have been published in the CEUR Workshop Proceedings \citep{Saracco2021,Petrocchi2022,Petrocchi2023,petrocchi2024overview,kruschwitz2025overview,pichel2026ceur}, which are freely accessible. Updated information on past and current ROMCIR editions can be found on the official \textit{Website}: \url{https://romcir.disco.unimib.it/}. The Website also features additional materials, such as the slides from keynote speeches delivered during the various editions, as well as a series of ``ROMCIR family photos'' that we are pleased to share.

\section{Workshop Organization}

The organization of the workshop, which enabled its successful realization, relied on the efforts of four main organizers, a Proceedings and Publicity Chair, and the members of the Program Committee, whom we would like to sincerely thank.

\paragraph{Marcos Fernández-Pichel.}
\begin{wrapfigure}{l}{0.213\textwidth}
\vspace{-13pt}
    \includegraphics[width=0.213\textwidth]{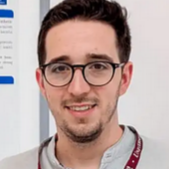}
      \vspace{-22pt}
\end{wrapfigure} He is an Assistant Professor at the University of Santiago de Compostela, Spain. He is also a research collaborator at Centro Singular de Investigación en Tecnoloxías Intelixentes (CiTIUS). He obtained his PhD in Computer Science (with honors) in 2023.
His research focuses on assessing the credibility of online health information, IR, and applying NLP to mental health. In 2022, the Royal Galician Academy of Sciences awarded him the Best Young Researcher Paper Award for his work published in Engineering Applications of Artificial Intelligence (EAAI). His work has been published in top-tier conferences (e.g., SIGIR, ECIR, EMNLP, EACL) and cross-interdisciplinary journals (NPJ Digital Medicine, NLE, EAAI, IEEE Transactions on Affective Computing).

\paragraph{Marinella Petrocchi.}
\begin{wrapfigure}{l}{0.213\textwidth}
    \includegraphics[width=0.213\textwidth]{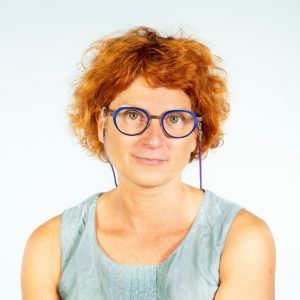}
      \vspace{-22pt}
\end{wrapfigure} She is a Senior Researcher at the Institute of Informatics and Telematics of the National Research Council (IIT-CNR) in Pisa, Italy, under the Trust, Security and Privacy research unit. She also collaborates with the Sysma unit at IMT School for Advanced Studies, in Lucca, Italy.
Her field of research lies between Cybersecurity, Artificial Intelligence, and Data Science. Specifically, she studies novel techniques for online fake news/fake accounts detection and automated methods to rank the reputability of online news media.
She is the author of several international publications on these topics, and she usually gives talks and lectures on the topic. She is the CNR Lead of the project Humane: Holistic sUpports to inforMAtioN disordEr, under the NRRP MUR program funded by the EU – NGEU.

\paragraph{Kevin Roitero.}
\begin{wrapfigure}{l}{0.213\textwidth}
\vspace{-13pt}
    \includegraphics[width=0.213\textwidth]{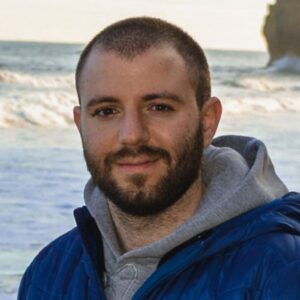}
      \vspace{-22pt}
\end{wrapfigure} He is a Tenure-Track Assistant Professor (RTDb) at the University of Udine, Italy. His research focuses on Artificial Intelligence, Natural Language Processing, Information Retrieval, and Crowdsourcing. He has collaborated with leading academic institutions such as the University of Sheffield, RMIT University, the University of Queensland, the Pioneer Center for Artificial Intelligence, and the University of Copenhagen, as well as with industry partners including Spotify.
His work has been published in top-tier conferences (e.g., SIGIR, WSDM, WWW, CIKM, HCOMP) and journals (TDKE, JDIQ, IRJ, IP\&M), and has earned him several awards, including the “con.Scienze2020” prize for the best Ph.D. thesis in Computer Science in Italy, multiple Best Paper Awards, and selection for the prestigious Heidelberg Laureate Forum, recognizing the top 200 young researchers in mathematics and computer science globally.

\paragraph{Marco Viviani.}
\begin{wrapfigure}{l}{0.213\textwidth}
\vspace{-13pt}
    \includegraphics[width=0.213\textwidth]{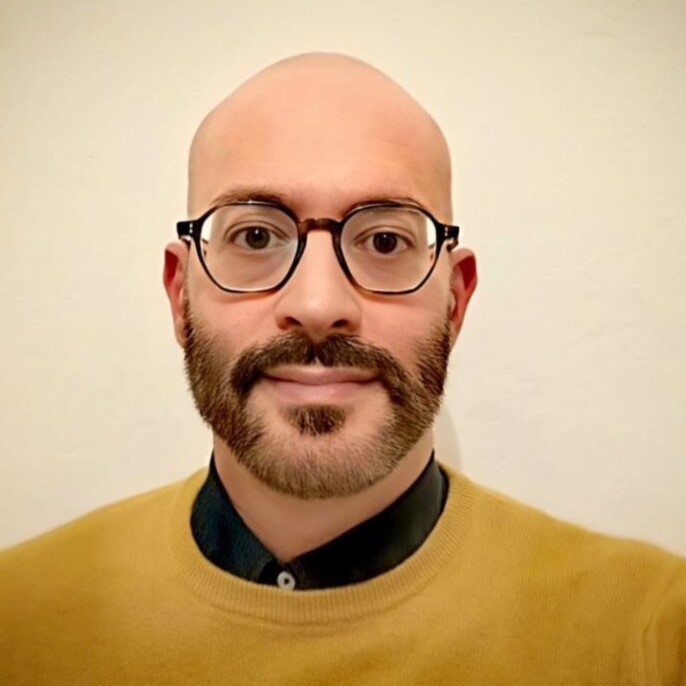}
    \vspace{-22pt}
\end{wrapfigure} He is an Associate Professor at the University of Milano-Bicocca, Department of Informatics, Systems, and Communication (DISCo). He works in the Information and Knowledge Representation, Retrieval and Reasoning (IKR3) Lab. He has been co-chair of several special tracks and workshops at international conferences, and general co-chair of MDAI 2019.
He is an Associate Editor of Social Network Analysis and Mining, Editorial Board Member of Online Social Networks and Media, and Guest Editor of several Special Issues in International Journals related to information disorder detection.
His main research activities include Social Computing, Information Retrieval, Text Mining, Natural Language Processing, Trust and Reputation Management, and User Modeling. On these topics, he has written several international publications.

\subsection*{Proceedings and Publicity Chair}

\begin{itemize}
    \item \textbf{Riccardo Lunardi}, University of Udine, Italy
\end{itemize}

\subsection*{Program Committee Members}
\begin{itemize}
    \item \textbf{Mario Ezra Aragon}, University of Santiago de Compostela, Spain
    \item \textbf{Marco Casavantes}, University of Santiago de Compostela, Spain
    \item \textbf{Manuel Couto-Pintos}, University of Santiago de Compostela, Spain
    \item \textbf{Gregor Donabauer}, University of Regensburg, Germany
    \item \textbf{David Elsweiler}, University of Regensburg, Germany
    \item \textbf{Carlos A. Iglesias}, Technical University of Madrid, Spain
    \item \textbf{Ema Kahr}, Vienna University of Economics and Business, Austria
    \item \textbf{Udo Kruschwitz}, University of Regensburg, Germany
    \item \textbf{David La Barbera}, University of Milano-Bicocca, Italy
    \item \textbf{David Losada}, University of Santiago de Compostela, Spain
    \item \textbf{Marcelo Mendoza}, Pontifical Catholic University of Chile, Chile
    \item \textbf{Selina Meyer}, University of Technology Nuremberg, Germany
    \item \textbf{Manuel Montes-Y-Gómez}, National Institute of Astrophysics, Optics and Electronics, Mexico
    \item \textbf{Anxo Pérez}, University of A Coruña, Spain
    \item \textbf{Daisy Romanini}, IIT-CNR, Italy
    \item \textbf{Alessandro Sapienza}, ISTC-CNR, Italy
    \item \textbf{Angelo Spognardi}, Sapienza University of Rome, Italy
    \item \textbf{Irene Sánchez Rodríguez}, IMT School for Advanced Studies Lucca, Italy
\end{itemize}

\begin{acknowledgments}
The ROMCIR 2026 Workshop was partially supported by  project re-DESIRE: \emph{DissEmination of ScIentific REsults} 2.0, funded by IIT--CNR; by SERICS (PE00000014) under the NRRP MUR program funded by  \#NGEU; by the NRRP ICSC National Research Centre for High-Performance Computing, Big Data and Quantum Computing (CN00000013), under the NRRP MUR program funded by \#NGEU; by project KURAMi: \textit{Knowledge-based, explainable User empowerment in Releasing private data and Assessing Misinformation in online environments} (Italian Ministry of University and Research -- PRIN 2022: 20225WTRFN), URL: \url{https://kurami.disco.unimib.it/}. Marcos Fernández-Pichel also thanks the financial support from the Agencia Estatal de Investigación (Spain) (PID2022-137061OB-C22 funded by MICIU/AEI/10.13039/ 501100011033), the Xunta de Galicia - Conselleria de Educación, Ciencia, Universidades e Formación Profesional (Centro de investigación de Galicia acreditación 2024-2027 ED431G-2023/04 and Reference Competitive Group accreditation ED431C 2022/19) and the European Union (European Regional Development Fund - ERDF).
\end{acknowledgments}

\section*{Declaration on Generative AI}
  The authors have not employed any Generative AI tools.
  

\bibliography{sample-ceur}




\end{document}